\documentclass[aps,superscriptaddress,twocolumn,showpacs]{revtex4-1}
\usepackage{amsmath,amssymb,graphicx,epsfig,color,bbm}
\usepackage[pass]{geometry}
\usepackage[colorlinks=true,citecolor=blue,linkcolor=blue,urlcolor=red,anchorcolor=black,filecolor=black]{hyperref}
\usepackage[resetlabels]{multibib}

\def\nn{\nonumber}

\newcommand{\ba}{\begin{eqnarray}}
\newcommand{\ea}{\end{eqnarray}}

\begin{document}

\title{Type-II Weyl cone transitions in driven semimetals}

\author{Ching-Kit Chan}
\affiliation{Department of Physics, Massachusetts Institute of Technology, Cambridge, Massachusetts 02139, USA}
\author{Yun-Tak Oh}
\affiliation{Department of Physics, Sungkyunkwan University, Suwon, 16419, Korea}
\author{Jung Hoon Han}
\affiliation{Department of Physics, Sungkyunkwan University, Suwon, 16419, Korea}
\author{Patrick A. Lee}
\affiliation{Department of Physics, Massachusetts Institute of Technology, Cambridge, Massachusetts 02139, USA}

\date{\today}

\begin{abstract}
Periodically driven systems provide tunable platforms to realize interesting Floquet topological phases and phase transitions. In electronic systems with Weyl dispersions, the band crossings are topologically protected even in the presence of time-periodic perturbations. This robustness permits various routes to shift and tilt the Weyl spectra in the momentum and energy space using circularly polarized light of sufficient intensity. We show that type-II Weyl fermions, in which the Weyl dispersions are tilted with the appearance of pocket-like Fermi surfaces, can be induced in driven Dirac semimetals and line node semimetals. Under a circularly polarized drive, both semimemtal systems immediately generate Weyl node pairs whose types can be further controlled by the driving amplitude and direction. The resultant phase diagrams demonstrate experimental feasibilities.
\end{abstract}
\pacs{71.20.-b, 03.65.Vf, 72.40.+w}
%73.43.-f(QHE), 03.65.Vf(topological phases), 72.40.+w(Photovoltaic effect), 71.20.-b(band structure)

\maketitle

\begin{figure}[t]
\begin{center}
\includegraphics[angle=0, width=1\columnwidth]{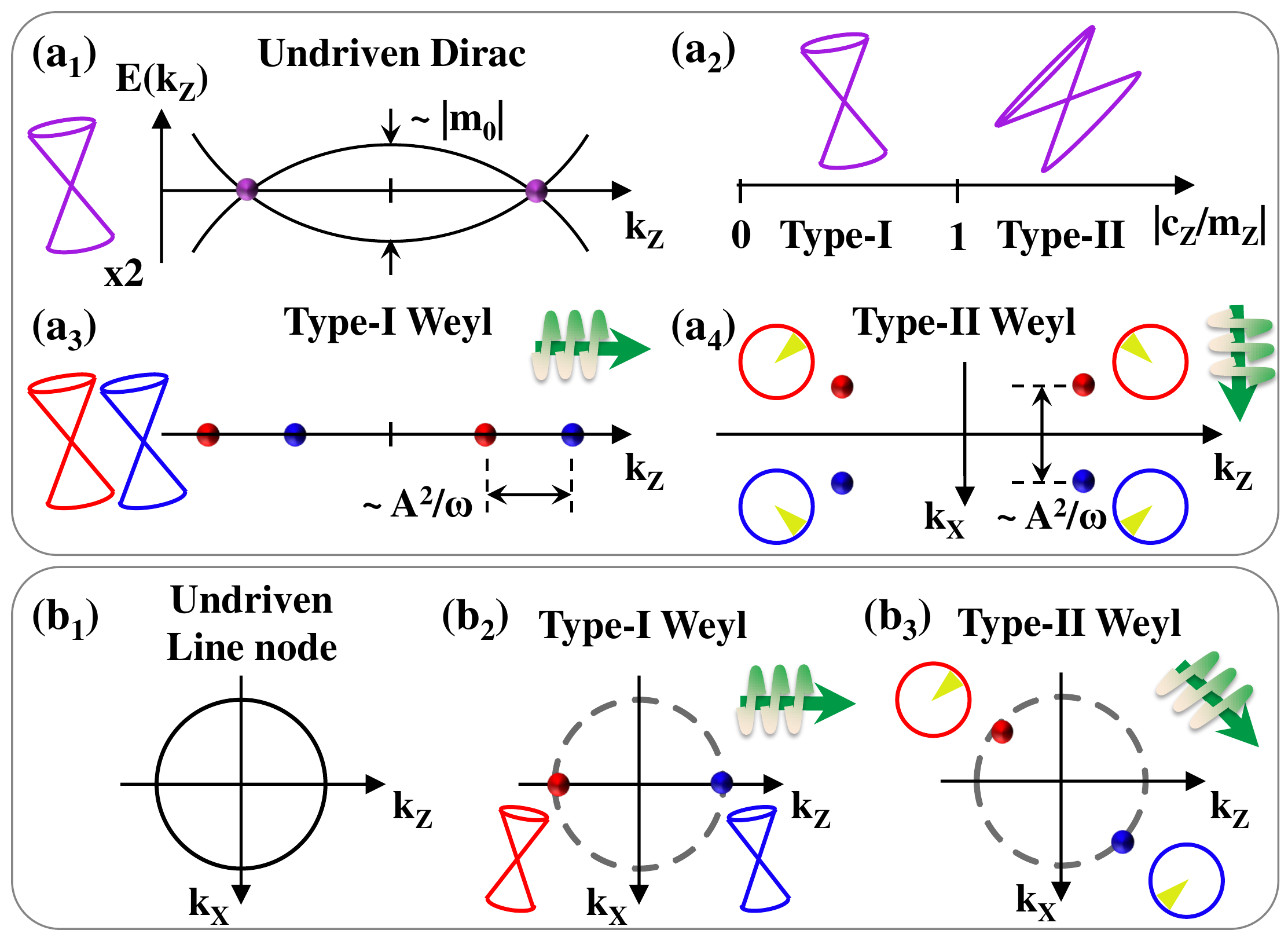}
\caption{Schematics of driven semimetals. ($\rm a_1$) Dispersion of a undriven DSM with two pairs of Dirac nodes. The energy gap at $\vec k=0$ is parameterized by $|m_0|$. ($\rm a_2$) Phase diagram of the undriven DSM. The type-I and type-II phases are determined only by a microscopic parameter $|c_z/m_z|$ [Eq.~(\ref{eq_undriven_HD})]. ($\rm a_3, a_4$) Driven DSMs. Green arrows denote driving directions. ($\rm a_3$) A $k_z$-drive splits the Dirac nodes into pairs of Weyl nodes along $k_z$, but cannot change the type of the cone. ($\rm a_4$) Driving along $k_x$ moves the nodes away from the $k_z$-axis and can induce a type-II Weyl transition. The tilting axis is related to the chirality. The circles show top views of the cones and the shaded regions mark the electron pockets. ($\rm b_1$) Undriven LNSM with a nodal ring. The dispersion in the $k_x-k_z$ plane can be generated by rotating that of ($\rm a_1$). An external drive gaps out the nodal line and creates a Weyl pair along the drive. Depending on model parameters, the photoinduced Weyl cone can belong to type-I ($\rm b_2$) or type-II ($\rm b_3$).}
\label{fig_schematic}
\end{center}
\end{figure}

%%%%%%%%%%%%%%%%%%%%%%%%%%%%%%%%%%%%%%%%%%%%%%%%%%%%%%%%%%%%%%%%%%%%%%%%%%%%%%%%%%%%%%%%%%%%%%%%%%%%%%%%%%%%%%%%%%%%%
\textit{Introduction.}---There has been rising interests in realizing Weyl fermions as 3D linearly band touching points in condensed matter systems \cite{PhysRevB.83.205101,Huang2015,PhysRevX.5.011029,Xu613,PhysRevX.5.031013}. The gapless, chiral and topological characters of Weyl points bring in a variety of exotic properties such as Fermi arc surface states, chiral anomaly and various quantum transport effects \cite{vafek14,0953-8984-27-11-113201}. These excitements are further elevated by a recent proposal of the so called ``type-II" Weyl fermion \cite{Soluyanov2015}, which is characterized by a sign change of the dispersion slope along some direction [Fig.~\ref{fig_schematic}($\rm a_2$)]. The type-II Weyl nodes happen at crossing points between electron and hole pockets in the band structure and are unique in solid state systems. They lead to many interesting features including open Fermi surfaces~\cite{Soluyanov2015}, anisotropic chiral anomaly \cite{Soluyanov2015,2016arXiv160404030Y,2016arXiv160408457U}, quantum Hall signatures \cite{2016arXiv160100890Z}, and more \cite{PhysRevB.91.115135,PhysRevLett.116.116803,2016arXiv160401028O,2016arXiv160403096M,2016arXiv160502671W}. There has been much theoretical effort to search for systems possessing type-II Weyl spectra \cite{2015arXiv151107440W,Chang2016,2016arXiv160304323K,2016arXiv160304624A,2016arXiv160401398M,2016arXiv160402124C,2016arXiv160503903L,PhysRevLett.115.265304} and experimental tests on the proposed Weyl semimetal (WSM) candidates have shown promising results \cite{2015arXiv150704847B,2016arXiv160306482H,2016arXiv160307318X,2016arXiv160308508D,2016arXiv160400139J,2016arXiv160401706L,2016arXiv160402116X,2016arXiv160402411B,2016arXiv160404218W,2016arXiv160405176W,2016arXiv160407079B,2016arXiv160408228T,2016arXiv160503380K}.

Given the dramatic changes of properties between the two types of Weyl fermion phases, it would be interesting to establish a tunable system with transitions between them. Recent advances in optical pump-probe experiments provide a reliable platform to manipulate Floquet-Bloch bands in laser-driven systems \cite{Wang453,Mahmood2016}. For example, an energy gap can be opened in the 2D Dirac spectrum of graphene by a circularly polarized drive~\cite{PhysRevB.79.081406,Sentef2015}. On the other hand, one can create various light-induced topological phases, such as Floquet topological insulators~\cite{Lindner2011} and Floquet Weyl phases~\cite{0295-5075-105-1-17004,PhysRevB.91.205445,PhysRevB.93.144114,PhysRevE.93.022209}, and transitions from Dirac to Weyl phases~\cite{PhysRevB.93.155107,2016arXiv160403399H}. Inspired by the high tunability of driven systems, in this paper, we examine the possibility of controllable transitions between the type-I and type-II Weyl fermion phases by laser-driving two classes of 3D semimetals: Dirac semimetals (DSM) and line node semimetals (LNSM) \cite{PhysRevLett.115.036806,PhysRevB.92.045108,PhysRevLett.115.036807}. Our work is different from previous Floquet Weyl phase studies in which the effect of tilt and the type-II Weyl phase were not considered. Naively, one would not anticipate such a transition, since a driven Weyl point can only be shifted perturbatively in the linear $\vec k \cdot \vec p$ regime \cite{chan16}. Surprisingly, we shall see below that by going beyond the linear theory, one can in principle photoinduce a type-I to type-II Weyl transition.

Our setups are schematically represented in Fig.~\ref{fig_schematic}. In a DSM respecting both the time-reversal (TR) and inversion (I) symmetries, there exist two TR-pairs of Dirac nodes separated by some momentum [Fig.~\ref{fig_schematic}($\rm a_1$)]. Due to the topological monopole characters of the nodes, an external drive cannot gap out the crossing, but only shift them \cite{chan16} depending on the chirality and the drive direction [Fig.~\ref{fig_schematic}($\rm a_3,\rm a_4$)]. In particular, when the Poynting vector of the drive is perpendicular to the separation momentum, the drive renormalizes the tilting and splitting energies differently and can give rise to a type-II Weyl transition [Fig.~\ref{fig_schematic}($\rm a_4$)]. On the other hand, a LNSM consists of a ring of degenerate states [Fig.~\ref{fig_schematic}($\rm b_1$)]. Upon driving, the nodal line is gapped, leaving a single pair of Weyl nodes and the type is determined by the drive direction [Fig.~\ref{fig_schematic}($\rm b_2$,$\rm b_3$)]. In both cases, anisotropy is crucial to realize the type-II transition. We note that these anisotropic parameters are commonplace in realistic materials \cite{PhysRevB.85.195320,Du2015}. While it is possible to realize the type-II phase in both systems, our results indicate that the transition effect on LNSM is experimentally more accessible. In the following, we will study the phase diagrams of driven DSM and LNSM separately. The corresponding photoinduced Hall effects across the Weyl transitions will also be discussed.

%%%%%%%%%%%%%%%%%%%%%%%%%%%%%%%%%%%%%%%%%%%%%%%%%%%%%%%%%%%%%%%%%%%%%%%%%%%%%%%%%%%%%%%%%%%%%%%%%%%%%%%%%%%%%%%%%%%%%%
\textit{Driving DSM.}---The low-energy Hamiltonian for a 3D Dirac semimetal with TR and I symmetries is $H_D = \sum_{\chi=\pm 1} H_{D,\chi}$ \cite{PhysRevB.85.195320,PhysRevB.88.125427,2016arXiv160502145J}, where
\begin{eqnarray}
\label{eq_undriven_HD}
H_{D,\chi} = c_i k_i^2 \sigma_0 + \left(m_0 - m_i k_i^2 \right) \sigma_z + v_x \chi k_x \sigma_x + v_y k_y \sigma_y. \ \ \ \
\end{eqnarray}
We have $m_{0,x,y,z}< 0$ to satisfy the band inversion condition. The Dirac nodes occur at $\vec k_0 = (0,0,\pm \sqrt{m_0/m_z})$ and the chirality is determined by $\text{sgn}(\chi k_{0,z})$. Existence of crystal symmetry is assumed to avoid Weyl nodes with opposite chirality ($\chi=\pm 1$) from gapping out each other \cite{PhysRevB.85.195320,PhysRevB.88.125427}.

Linearizing around the Dirac node, we have the energy dispersion:
\begin{eqnarray}
\label{eq_undriven_ED}
&&E_{D,\chi} (\vec q = \vec k - \vec k_0) = T(\vec q) \pm U(\vec q) \\
&&\ \ \ = 2 k_{0,z} (c_z q_z)\pm \left[4 k_{0,z}^2( m_z q_z)^2 + (v_x q_x)^2 + (v_y q_y)^2 \right]^{1/2}. \nn
\end{eqnarray}
The first T-term characterizes the tilting of the Dirac cone along the $k_z$ direction, and the second U-term describes the splitting of the cone. The node becomes type-II when the tilting energy dominates, i.e. when there exists a direction ($\theta_c,\phi_c$) such that $|T(q,\theta_c,\phi_c)| \geq |U(q,\theta_c,\phi_c)|$ \cite{Soluyanov2015}. All four Weyl nodes have the same tilting amplitude.

For the undriven system, the phase diagram is simply determined by the ratio $|c_z/m_z|$ [Fig.~\ref{fig_schematic}($\rm a_2$)]: type-I when $|c_z/m_z|<1$ and type-II when $|c_z/m_z|> 1$. Other parameters can only affect the amount of tilt and are not relevant to the transition. Below, we consider a type-I system with $|c_z/m_z|<1$ and study the effect of a circularly polarized light on it. We separately discuss the consequences for driving along $k_z$ and $k_x$ directions.

In the presence of a $k_z$-drive (i.e. Poynting vector along $k_z$), we have $\vec k \rightarrow \vec k + A (\cos \omega t, \xi \sin \omega t, 0)$, where $\xi$ represents the light polarization. In the high frequency regime \cite{doi:10.1080/00018732.2015.1055918}, the Floquet Hamiltonian gains additional contributions:
\begin{eqnarray}
\Delta H^F_{D,\chi} &=& -\left( \frac{v_x v_y \chi \xi}{\omega} + \frac{m_x+m_y}{2} \right) A^2 \sigma_z  \\
&&\ - \left( \frac{2 m_x v_y \xi k_x A^2}{\omega} \right) \sigma_x - \left( \frac{2 m_y v_x \chi \xi k_y A^2}{\omega} \right) \sigma_y. \nn
\end{eqnarray}
This is equivalent to a shift of $v_{x(y)} \rightarrow v_{x(y)} -2 m_{x(y)} v_{y(x)} \chi \xi A^2/\omega$ and $m_0 \rightarrow m_0 - \left( v_x v_y \chi \xi/\omega + (m_x+m_y)/2 \right) A^2$ for the undriven Hamiltonian. The subsequence of the mass shift is to split each Dirac cone into two Weyl nodes along the $k_z$-direction [Fig.~\ref{fig_schematic}($\rm a_3$)]. However, as discussed above, these parameter shifts are irrelevant to the Weyl transition.

The situation dffers when we drive along $k_x$. The Floquet corrections are:
\begin{eqnarray}
\label{eq_driven_HDF_kx}
\Delta H^F_{D,\chi} = -\left(\frac{m_y+m_z}{2}\right) A^2 \sigma_z + \left( \frac{2 m_z v_y \xi  k_z A^2}{\omega} \right)\sigma_x.\ \ \ \
\end{eqnarray}
Together with Eq.~(\ref{eq_undriven_HD}), the Weyl nodes are split along the $k_x$ direction: $\vec k_0' = \left( \lambda_{D} k_{0,z}',0,k_{0,z}'\right)$ with a characteristic coupling:
\begin{eqnarray}
\label{eq_lambdaD}
\lambda_{D} =- \frac{2 m_z v_y \chi \xi  A^2}{v_x \omega},
\end{eqnarray}
and $k_{0,z}'= \pm \{(m_0 -\frac{m_y+m_z}{2}A^2)/(m_z + m_x \lambda_D^2)\}^{1/2}$. This chirality-dependent node shift along the driving direction is again expected [Fig.~\ref{fig_schematic}($\rm a_4$)].

We now examine the possibility of type-II transition. Our result shows that the tilting energy is rotated from $T(\vec q)\propto q_z$ to $T(\vec q) \propto \left(c_z q_z + c_x \lambda_D q_x\right)$ which is chirality-dependent, and the transition is no longer simply governed by the single parameter $|c_z/m_z|$. The effect becomes more transparent by rotating the $q_x-q_z$ plane: $\vec q' = R\left[\theta_q = \tan^{-1}\left(-c_x \lambda_{D}/c_z\right)\right] \vec q$, after which we have $T(\vec q')\propto q_z'$. Along $\vec q_z'$,
\begin{eqnarray}
\label{eq_tiltingfactor}
\left|\frac{T(\hat q_z')}{U(\hat q_z')} \right| = \left| \frac{c_z}{m_z} \right| \frac{1+ \Delta c^2 \lambda_{D}^2}{ \sqrt{(1+ \Delta c \Delta m \lambda_{D}^2)^2 + F(v_x,k_{0,z})^2} }, \ \
\end{eqnarray}
where $\Delta c = c_x/c_z$ and $\Delta m = m_x/m_z$ represent the anisotropy, and $F(v_x,k_{0,z})\propto v_x/k_{0,z}$. The bare tilting ratio $|c_z/m_z|$ is now renormalized by a factor that depends on the driving intensity and anisotropy. Particularly, a type-I Dirac node can be driven to a type-II Weyl node as long as the renormalization exceeds $|c_z/m_z|^{-1}$. If we further assume $\Delta m\sim 1$ and a negligible $v_{x}$, the effect will be most prominent when $\Delta c \sim -1/\lambda_D^2$, i.e. in the anisotropic regime. Physically, the transition mechanism is due to the different renormalizations of $T(\vec q)$  and $U(\vec q)$ by the drive. We note that both the photoinduced node shift and tilt are caused by $A^2$ terms in the Floquet Hamiltonian, while earlier works tend to focus on the shift effect only~\cite{PhysRevB.93.155107,2016arXiv160403399H,chan16}.

%Note that Eq.~(\ref{eq_tiltingfactor}) only gives a necessary condition for the transition.

\begin{figure}[tb]
\begin{center}
\includegraphics[angle=0, width=1\columnwidth]{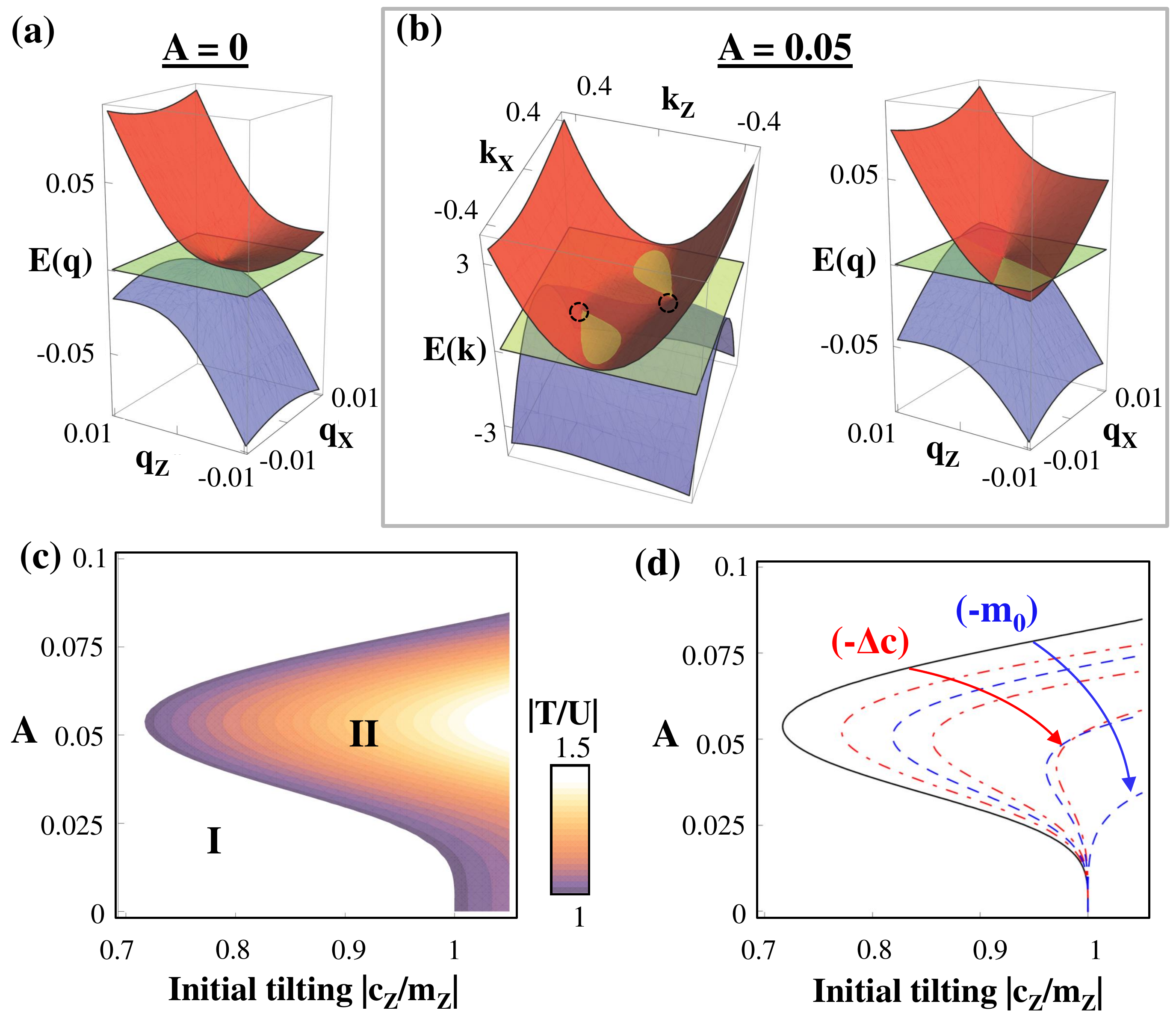}
\caption{Phases and phase diagrams of driven DSMs. (a) Undriven type-I Dirac dispersion. (b) Dispersion when driven along $k_x$. Left panel: the Dirac node is split into a type-II Weyl pairs with electron and hole pockets. Only the $\chi=1$ nodes are shown. The $\chi=-1$ counterparts can be obtained by reversing the $k_x$ axis. Right panel: zoom-in dispersion in the vicinity of one of the split nodes. (c) Phase diagram of the driven DSM showing the photoinduced type-II Weyl phase (A in unit of $\rm \AA^{-1}$). The false color scale gives the maximum tilting ratio $\rm |T/U|$ in the type-II phase and the phase boundary happens at $\rm max\left(|T/U|\right)=1$. In (a-c) $-m_0 =\rm 0.5~eV$ and $-\Delta c=1$, where in (a-b) $c_z/|m_z|=0.8$. The mass and anisotropy are important to induce the type-II phase. (d) Shrinking of the type-II region as we reduce $-m_0$ (dashed blue, from $\rm 0.5~eV$ to $\rm 0.37,0.23,0.1~eV$) and $-\Delta c$ (dot-dashed red, from $1$ to $0.8,0.5,0$), respectively. Common parameters: $v_x=v_y=\rm 2.5~eV \AA$, $m_{x,y,z}=\rm -10~eV \AA^2$, $c_x=c_y$, $\omega=\rm 0.1~eV$ and $\chi=\xi=1$. }
\label{fig_Dirac}
\end{center}
\end{figure}

%In $\rm Na_3 Bi$, the anisotropy is large $(-\Delta c \sim 1)$ but the mass is small $(\rm -m_0\sim 0.1~eV)$~\cite{PhysRevB.85.195320}.

Figure~\ref{fig_Dirac} details the phase diagrams obtained from the Floquet dispersion. Before the drive, the type-I and type-II phases are separated at $|c_z/m_z|=1$. With some initial tilting, the drive causes a transition from the type-I to type-II Weyl phase [Fig.~\ref{fig_Dirac}(c)]. Two essential quantities are responsible for such a transition: the anisotropy $\Delta c$ and a relatively large mass $|m_0|$. The mass requirement originates from the fact that a tiny mass gives a small $k_{0,z}$, which leads to a large $F$ value in Eq.~(\ref{eq_tiltingfactor}) and thus obstructs the tilt. Figure~\ref{fig_Dirac}(d) plots how the type-II phase region shrinks with the decrease of the mass and anisotropy. We remark that further increment of the drive can gap out all four nodes at $A_{\rm gap}\sim \{m_0/m_{x,y,z}\}^{1/2}$ because of the sign change of the renormalized mass. The choice of parameters used in Fig.~\ref{fig_Dirac} are mostly based on realistic DSM materials~\cite{note2}.

In order to experimentally access the type-II Weyl transition, a DSM material with some anisotropy ($\rm \Delta c \neq 1$), finite initial tilts ($\rm |c_z/m_z| \neq 0$) and a large $\rm |m_0|$ (or equivalently small $\rm v_{x,y}$) is desirable. We have examined the driven phase diagrams of $\rm Na_3 Bi$ \cite{PhysRevB.85.195320} and $\rm Ba Au Bi$ \cite{Du2015} and found that $\rm Ba Au Bi$ is a suitable DSM candidate to observe the photoinduced type-II Weyl phase~\cite{note2}. The reason is that even though both materials have finite anisotropy and initial tilts, the Dirac cone in $\rm Na_3 Bi$ has a small mass ($\rm m_0=-0.087~eV$) as compared to $\rm Ba Au Bi$ ($\rm m_0=-0.22~eV$), and the tilt renormalization in $\rm Na_3 Bi$ is suppressed as mentioned above.

%%%%%%%%%%%%%%%%%%%%%%%%%%%%%%%%%%%%%%%%%%%%%%%%%%%%%%%%%%%%%%%%%%%%%%%%%%%%%%%%%%%%%%%%%%%%%%%%%%%%%%%%%%%%%%%%%%%%%%
\textit{Driving LNSM.}---Driving a LNSM can also create type-II Weyl nodes with much relaxed conditions. A LNSM involves a ring of band touching states and its low-energy Hamiltonian takes the form \cite{PhysRevLett.115.036806,PhysRevB.92.045108,PhysRevLett.115.036807}:
\begin{eqnarray}
H_{LN} = \left(c_0 + c_i k_i^2 \right) \sigma_0 + \left(m_0 - m_i k_i^2 \right) \sigma_z + v_y k_y \sigma_y,\ \
\label{eq_undriven_HLN}
\end{eqnarray}
which is the zero $v_x$ limit of the Dirac Hamiltonian [Eq.~(\ref{eq_undriven_HD})]. This Hamiltonian can be constructed by imposing both TR and I symmetries on a spinless system \cite{PhysRevLett.115.036806,PhysRevB.92.045108,PhysRevLett.115.036807}. Under the same band inversion condition $m_{0,x,y,z} < 0$, a line-node happens along an ellipse defined by $m_x k_x^2 + m_z k_z^2 =m_0$ and $k_y=0$. Interesting features of LNSMs include a $\pi$-Berry phase along any loop pierced by the LN and nearly-flat surface states \cite{PhysRevB.84.235126}. LNSMs have been experimentally observed by photoemissions recently \cite{2015arXiv150900861S,2016arXiv160300934W,2016arXiv160400720N}.

\begin{figure}[tb]
\begin{center}
\includegraphics[angle=0, width=1\columnwidth]{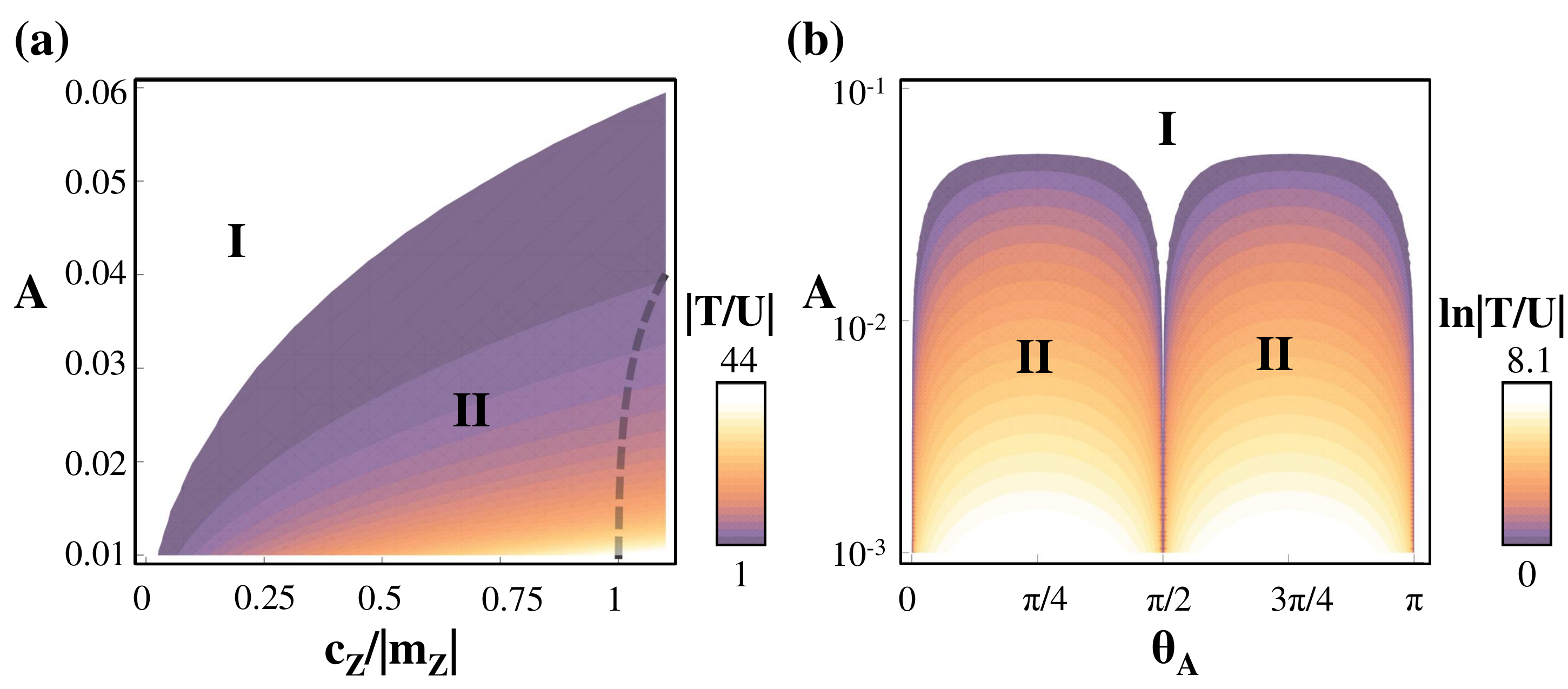}
\caption{Phase diagrams of driven LNSMs using a realistic and small mass $-m_0=\rm 0.1~eV$. The induced type-II phase region is much larger. (a) Phase diagram at a fixed driving angle $\theta_A=\pi/4$. The corresponding phase boundary for the DSM case is marked by the dashed line for comparison. (b) Phase diagram at fixed $c_z/|m_z|=0.8$. Different from the DSM case, the type-II phase can occur even for small driving amplitudes in LNSMs. The requirements on the mass and drive to photoinduce a type-II phase in LNSMs are much relaxed. Common parameters: $v_y=\rm 2.5~eV \AA$, $m_{x,y,z}=\rm -10~eV \AA^2$, $c_x=c_y$, $\omega=\rm 0.1~eV$, $\xi=1$ and $-\Delta c=1$.}
\label{fig_Linenode}
\end{center}
\end{figure}

An external $k_z$-drive immediately gaps out the LN except at two points $\vec k_0 = ( 0,0,\pm \sqrt{m_0'/m_z})$, which are photoinduced Weyl nodes [Fig.~\ref{fig_schematic}($\rm b_2$)]. This can be seen from the Floquet terms:
\begin{eqnarray}
\Delta H^F_{LN} = -\left( \frac{m_x+m_y}{2} \right) A^2 \sigma_z - \left( \frac{2 m_x v_y \xi k_x A^2}{\omega} \right) \sigma_x,\ \ \ \
\end{eqnarray}
where $\xi$ controls the node chirality and $m_0'=\left[m_0-(m_x+m_y)A^2/2 \right]$. The driven LNSM now acquires an effective $v_x=-2 m_x v_y A^2/\omega \ll v_y$ and the emergence of the $\sigma_x$ term gaps out the nodal line. Since the overall Hamiltonian has the same form as Eq.~(\ref{eq_undriven_HD}) with $\xi$ playing the role of $\chi$, a pair of Weyl node is induced and its type is simply determined by the bare value $|c_z/m_z|$. As such, no Weyl type transition can be caused by increasing $A$. Another interesting consequence is that the 2D surface state enclosed by the original nodal ring is no longer supported by the bulk and is expected to transform to a Fermi arc connecting the Weyl pair.

Now, we rotate the drive direction on the $k_z-k_x$ plane with $\vec A(t) = A (\cos\theta_A \cos\omega t, \xi \cos\theta_A \sin \omega t, $ $ -\sin\theta_A \cos \omega t)$, where $\theta_A$ defines the driving angle away from the $k_z$-axis. The Floquet contribution becomes:
\begin{eqnarray}
\Delta H^F_{LN} &=& -\left(\frac{m_x \cos^2\theta_A + m_y + m_z \sin^2\theta_A}{2}\right) A^2\sigma_z \nn \\
&&-  \frac{2 v_y \xi A^2  }{\omega} (m_x k_x \cos\theta_A - m_z k_z \sin\theta_A)\sigma_x, \ \ \
\end{eqnarray}
which has the same form as the DSM Hamiltonian driven along $k_x$ [Eq.~(\ref{eq_driven_HDF_kx})], but with two significant distinctions: a photoinduced $v_x$ and a large coupling
\begin{eqnarray}
\label{eq_lambdaLN}
\lambda_{LN} = \frac{m_z}{m_x} \tan\theta_A
\end{eqnarray}
being controlled by $\theta_A$. The nodes are now shifted to $\vec k_0'=\left(\lambda_{LN} k_{0,z}',0,k_{0,z}'\right)$ [Fig.~\ref{fig_schematic}($\rm b_3$)] with $k_{0,z}'=\pm \{[m_0 -(m_x \cos^2\theta_A + m_y + m_z\sin^2\theta_A)A^2/2]/(m_z + m_x \lambda_{LN}^2)\}^{1/2}$. The type-II phase analysis just follows that of the driven DSM case. Notably, due to the small $v_x$ here, a large mass is no longer required. Moreover, since a large coupling with $\lambda_{LN} \sim O(1)$ can be achieved by tuning $\theta_A$, we anticipate a type-II Weyl transition by a weak drive.

Figure~\ref{fig_Linenode} illustrates the type-II Weyl phase transition for a driven LNSM. Parameter values similar to those in the DSM study are chosen for comparison purposes. As we rotate the drive on the $k_z-k_x$ plane, the Weyl pair undergoes transitions between the two types depending on parameter values. In sharp contrast to the DSM case, the photoinduced $v_x$ in LNSM guarantees a negligible $F$ contribution in Eq.~(\ref{eq_tiltingfactor}) and hence removes the large mass requirement. Figure~\ref{fig_Linenode}(a) contrasts the type-II phase regions in driven LNSM and DSM. The type-II phase remains in the LNSM case even when a realistically small mass is used. Besides, the type-II phase persists for weak drive amplitudes [Fig.~\ref{fig_Linenode}(b)]. With the less stringent demands on the materials and the drive, a type-II Weyl phase can be achieved more easily in LNSMs.

We remark that LNSM formed in a spinless system requires the absence of spin-orbit coupling (SOC). Existence of SOC can gap out the nodal line and result in DSM or WSM \cite{2016arXiv160304744W}. Yet, additional nonsymmorphic symmetries can protect the nodal line from being gapped \cite{PhysRevB.92.081201}. Even if there is no nonsymmorphic symmetry, SOC shall not be an issue here as long as it is weak, since the crucial requirement for the type-II transition is small $\rm v_x$, rather than zero $\rm v_x$.

%%%%%%%%%%%%%%%%%%%%%%%%%%%%%%%%%%%%%%%%%%%%%%%%%%%%%%%%%%%%%%%%%%%%%%%%%%%%%%%%%%%%%%%%%%%%%%%%%%%%%%%%%%%%%%%%%%%%%%
\textit{Anomalous Hall effects.}---Lastly, we discuss implications on the  anomalous Hall effect (AHE) as the Weyl fermion transits between the two types. An ideal type-I Weyl pair in a TR-breaking system can generate an anomalous Hall conductance $\sigma_{\alpha\beta}(\mu=0)=e^2\Delta k_\gamma/(2\pi h)$ determined by the momentum separation $\Delta k_\gamma$ \cite{PhysRevB.84.075129}. For a type-II Weyl pair, however, the presence of Fermi pockets modifies the Berry curvature contributions and thus should result in tilt-dependent Hall signals \cite{2016arXiv160100890Z}. Furthermore, the dependence on chemical potential is expected to be stronger in the type-II phase due to the imbalance between the electron and hole pockets away from the neutrality. Therefore, it was predicted that the AHE will experience sharp change across the phase transition~\cite{2016arXiv160100890Z}.

However, this anomalous Hall signature occurs only when the type-I phase consists of ideal Weyl pairs, i.e. when there is no other electron or hole pocket away from the Weyl nodes. For realistic Weyl materials, there can be Fermi pockets near the Weyl node in the type-I phase, in which the Hall conductivity is not simply related to the node separation. A transition to the type-II phase can be accomplished by joining the pockets with the band crossing, and in this case we do not anticipate dramatic transitional behaviors in the AHE across the Weyl phase transition~\cite{note2}.

%%%%%%%%%%%%%%%%%%%%%%%%%%%%%%%%%%%%%%%%%%%%%%%%%%%%%%%%%%%%%%%%%%%%%%%%%%%%%%%%%%%%%%%%%%%%%%%%%%%%%%%%%%%%%%%%%%%%%%
\textit{Conclusion.}---We have analyzed how photoinduced transitions between the type-I and type-II Weyl fermion phases can occur in driven DSMs and LNSMs. These schemes provide highly tunable routes to construct type-II Weyl fermions and study the transitions across. Anisotropic tilting are critical to allow the type-II phase and are common in realistic materials. The weak field requirement for the LNSM case shall facilitate future experimental implementations. Our general phase diagrams, in line with specific material studies, provide a guideline for existing and future materials to probe the transitional effects by time-resolved photoemission experiments. The resultant Floquet-Weyl phase also paves a way to explore quenching of Fermi surface and surface state topology.

\textit{Acknowledgements.}--- P.A.L. acknowledges the support from DOE Grant No. DE-FG02-03-ER46076 and the Simons Fellows Program. Y.-T.O. was supported by a Global Ph.D. Fellowship Program through the National Research Foundation of Korea (NRF) funded by the Ministry of Education (NRF-2014H1A2A1018320). P.A.L. and C.-K.C. thank the hospitality of the CMT group at Caltech where this work was carried out.

\textit{Note added.}--- During the completion of this manuscript, we became aware of a work about photoinducing Weyl points in driven LNSMs \cite{2016arXiv160504404Y}, but the type-II transition was not discussed.

\bibliography{Floquet-Weyl-II}

%%%%%%%%%%%%%%%%%%%%%%%%%%%%%%%%%%%%%%%%%%%%%%%%%%%%%%%%%%%%%%%%%%%%%%%%%%%%%%%%%%%%%%%%%%%%%%%%%%%%%%%%%%%%%%%%%%%%%%%%%%%%%%%%%%%%%%%%%%%%%%
\clearpage
\begin{widetext}

\begin{center}
\large{\bf Supplemental Material:\\ Type-II Weyl cone transitions in driven semimetals}\\
\vspace{14pt}
\normalsize{Ching-Kit Chan, Yun-Tak Oh, Jung Hoon Han, and Patrick A. Lee}
\end{center}
%\vspace{14pt}

%%%%%%%%%%%%%%%%%%%%%%%%%%%%%%%%%%%%%%%%%%%%%%%%%%%%%%%%%%%%%%%%%%%%%%%%%%%%%%%%%%%%%%%%%%%%%%%%%%%%%%%%%%%%%%%%%%%%%
\section{Phase diagrams for specific Dirac semimetals}

\begin{figure}[b]
\begin{center}
\includegraphics[angle=0, width=0.6\columnwidth]{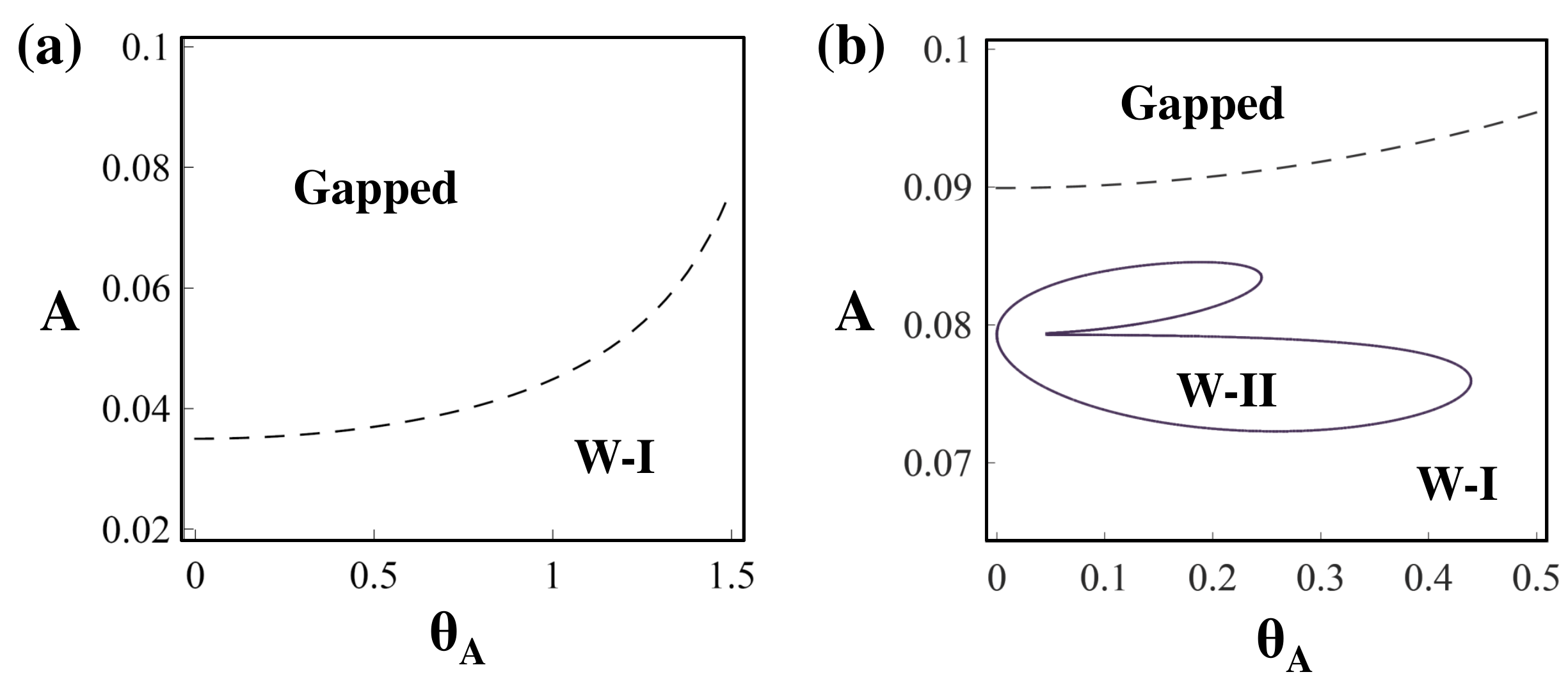}
\caption{Phase diagrams of driven (a) $\rm Na_3 Bi$ and (b) $\rm Ba Au Bi$ as a function of driving amplitude A and angle $\rm \theta_A$. Only the $\rm \chi=+1$ Weyl pairs are considered. Both systems transit from the Dirac phase to the type-I Weyl (W-I) phase in the presence of a circularly polarized drive. Only $\rm Ba Au Bi$ can further induce a type-II Weyl (W-II) transition because of its relatively larger mass $\rm |m_0|$ and smaller velocity v. A large field can gap out the Weyl nodes in both systems.}
\label{fig_compare_Dirac_supple}
\end{center}
\end{figure}

In this section, we provides specific studies of the driven phase diagrams of realistic Dirac semimetal materials. Based on previous band structure calculations, we examine $\rm Na_3 Bi$ \cite{PhysRevB.85.195320_s} and $\rm Ba Au Bi$ \cite{Du2015_s} whose low-energy effective Hamiltonian parameters are explicitly given in the literature. In both materials, the undriven Dirac cones have some anisotropy and finite tilts which are necessary for the type-II Weyl transition. However, only $\rm Ba Au Bi$ can support the type-II transition due to its relatively large mass $\rm m_0$ (or equivalently small $\rm v_{x,y}$), in agreement with the discussion in the main text.

Both $\rm Na_3 Bi$ and $\rm Ba Au Bi$ have time-reversal and inversion symmetries and their effective Hamiltonians $\rm H_{D,\chi=\pm 1}$ are given by Eq.~(1) in the main text, with $\rm v_x=v_y=v$, $\rm c_x=c_y$ and $\rm m_x=m_y$. Since the systems are cylindrically symmetric about the $\rm k_z$ axis, we consider a drive on the $\rm k_x-k_z$ plane such that the Poynting vector points along $(\sin\theta_A \hat k_x + \cos\theta_A \hat k_z )$. We have the vector potential $\vec A(t) = A(\cos\theta_A \cos\omega t, \xi \sin\omega t, -\sin\theta_A \cos\omega t)$. The effective Floquet Hamiltonian for the driven Dirac system becomes:
\begin{eqnarray}
\label{eq_HD_s}
H_{D,\chi}^F &=& c_i k_i^2 \sigma_0 + \left [m_0 - \frac{A^2}{2}\left(m_x \cos^2\theta_A +m_y + m_z \sin^2\theta_A \right) -\frac{A^2}{\omega}\chi\xi v_x v_y \cos\theta_A -m_i k_i^2 \right]\sigma_z \nn \\
&&+ \left[ v_x \chi k_x + \frac{2 A^2 \xi v_y}{\omega} \left(-m_x k_x \cos\theta_A +m_z k_z \sin\theta_A \right) \right] \sigma_x
  + \left[ v_y k_y + \frac{2 A^2 \chi \xi v_x}{\omega} \left(-m_y k_y \cos\theta_A \right) \right] \sigma_y.
\end{eqnarray}
Unimportant constant terms are dropped. Eq.~(3) and Eq.~(4) in the main text correspond to $\theta_A=0$ and $\theta_A=\pi/2$ of the above Hamiltonian, respectively.

Fig.~\ref{fig_compare_Dirac_supple} compares the phase diagrams computed from Eq.~(\ref{eq_HD_s}) for $\rm Na_3 Bi$ and $\rm Ba Au Bi$ as we scan over $\rm A $ and $\rm \theta_A$ with $\rm \omega=0.1~eV$ and $\rm \xi=-1$. In each system, the $\rm \chi=+1$ and $\rm \chi=-1$ pairs behave differently except at $\theta_A = \pi/2$. We only plot the phases for the $\rm \chi=+1$ pairs in Fig.~\ref{fig_compare_Dirac_supple}. The $\rm \chi=-1$ pairs belong to the type-I Weyl phase throughout the same phase region. The parameter values $\rm (c_x, c_z, m_0, m_x, m_z, v)$ are $\rm (-8.4~\AA^2, 8.75~\AA^2, -0.087, -10.36~\AA^2, -10.64~\AA^2, 2.46~\AA)~eV$ and $\rm (2.25~\AA^2, -0.34~\AA^2, -0.22, -7.95~\AA^2, -1.95~\AA^2, 1.37~\AA)~eV$ for $\rm Na_3 Bi$ \cite{PhysRevB.85.195320_s} and $\rm Ba Au Bi$ \cite{Du2015_s}, respectively. Both systems have some anisotropy ($\rm \Delta c = c_x/c_z \neq 1$) and finite initial tilts ($\rm |c_z/m_z| \neq 0$). When $\rm A=0$, both systems start with the Dirac phase. They are immediately driven to the type-I Weyl phase by a weak drive. However, only $\rm Ba Au Bi$ can undergo the type-I to type-II transition by increasing A. The difference between the two systems comes from the relatively larger (more negative) mass $\rm m_0$ and smaller velocity $\rm v$ in $\rm Ba Au Bi$, which are crucial for the photoinduced tilt renormalization as described in the main text.

A number of features of the phase diagrams are anticipated from the discussion in the main text. First, by further increasing A, the renormalized mass ($\rm m_0'=m_0 - A^2[m_x (\cos^2\theta_A +1)/2 + m_z \sin^2\theta_A ] -A^2\chi\xi v^2 \cos\theta_A/\omega$) changes sign for the $\rm \chi=+1$ pairs. The band inversion condition no longer holds and thus the Weyl nodes are gapped in both systems. Moreover, as mentioned in the main text, there can be no type-II transition due to a $\rm k_z$-drive ($\rm \theta_A=0$) in which there is no renormalization of the tilt. A type-II transition can only occur when $\rm \theta_A \neq 0$.

%%%%%%%%%%%%%%%%%%%%%%%%%%%%%%%%%%%%%%%%%%%%%%%%%%%%%%%%%%%%%%%%%%%%%%%%%%%%%%%%%%%%%%%%%%%%%%%%%%%%%%%%%%%%%%%%%%%%%%
\section{Anomalous Hall signatures}

Here we provide more detailed discussion about the quantum anomalous Hall effects as the Weyl dispersion transits across type-I and type-II. As mentioned in the main text, an ideal type-I Weyl pair gives a Hall conductance $\sigma_{\alpha\beta}(\mu=0)=e^2\Delta k_\gamma/(2\pi h)$ $\Delta k_\gamma$. For a type-II Weyl pair, this formula no longer holds. The Fermi pockets largely modifies the Berry curvature and hence the Hall signals. Because of this, we expect a stronger chemical potential dependence of $\sigma_{\alpha\beta}$ in the type-II phase due to the imbalance between the electron and hole pockets away from the neutrality.

Figure~\ref{fig_AHE_supple}($\rm a_1$) plots the computed anomalous Hall conductance for the driven LNSM as we move across the Weyl transition for a given $\theta_A$. We have assumed that $\mu$ is small enough so that the anomalous Hall effect is purely intrinsic~\cite{0953-8984-27-11-113201_s}. We first observe the expected behavior at $\mu=0$ that $\sigma_{xy}/\left[e^2/\left(2\pi h\right)\right]=\Delta k_{0,z} $ in the type-I phase and drops as the Weyl pair is driven into the type-II region. At finite $\mu$, even though the large electron and hole pockets contribute opposite Berry curvatures to $\sigma_{xy}$, the pocket sizes are asymmetric. The net outcome is the enhanced $\mu$ dependence of $\sigma_{xy}$ in the type-II phase.

Importantly, such a Hall signature applies only when the type-I phase consists of ideal Weyl pairs. In fact, there are two distinct routes for the Weyl type transitions. In the ideal case, the type-II cone simply tilts upright along with the disappearance of the pockets [Fig.~\ref{fig_AHE_supple}($\rm a_2$)]. Alternatively, the pockets connecting with the Weyl cone can detach and move away from the node [Fig.~\ref{fig_AHE_supple}(b)]. Since the pockets are still present in the latter case, the Hall conductance is not simply determined by the momentum separation as mentioned above. The routes of transition depend on microscopic details and both transition routes are possible in our driven system. Similar analyses also apply to the other transport behaviors.

\begin{figure}[tb]
\begin{center}
\includegraphics[angle=0, width=0.6\columnwidth]{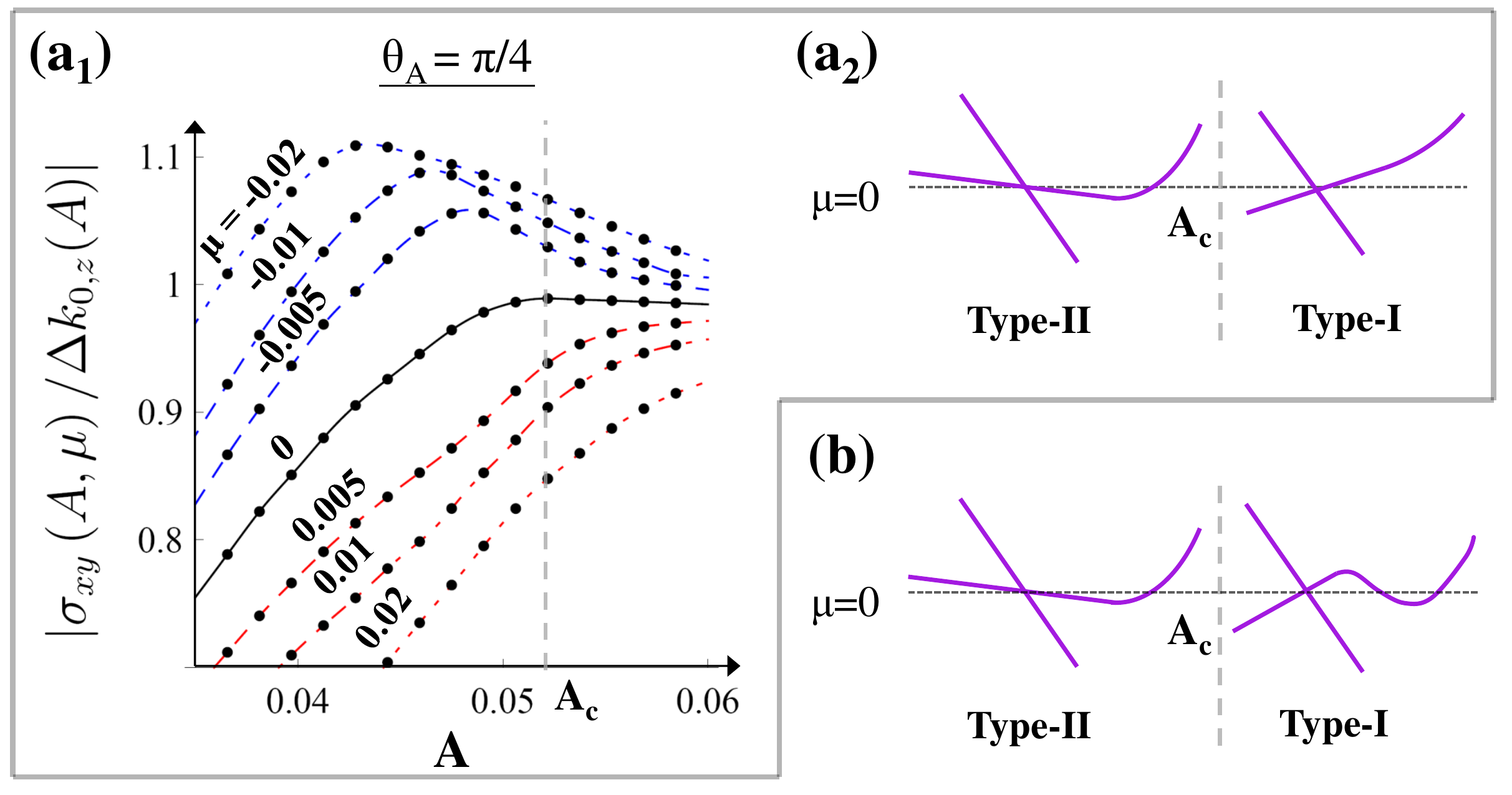}
\caption{($\rm a_1$) Hall conductances as a function of $A$ for different $\mu$ in the driven LNSM (in unit of $e^2/(2\pi h)$). The vertical dashed line denotes the phase boundary. Same parameters as in Fig.~3(b) in the main text at $\theta_A=\pi/4$. In the type-II regime $A<A_c$, the $\mu$ dependence is enhanced due to the presence of Fermi pockets. ($\rm a_2$) Dispersion transition from a type-II to an ideal type-I Weyl phase, leading to transitional behaviors of $\sigma_{xy}$. (b) Another scenario where the pockets remain across the transition.}
\label{fig_AHE_supple}
\end{center}
\end{figure}

\end{widetext}
\end{document}